\documentclass[12pt]{iopart}

\usepackage{graphicx}
\newcommand{\mF}{f}
\newcommand{\G}{\gamma}

\newcommand{\p}{(\partial_xh)}
\newcommand{\pd}{(\partial_x^2h)}
\newcommand{\ppt}{(\partial_x^3h)}
\newcommand{\pc}{(\partial_x^4h)}

\begin{document}

\title[Generic equations for pattern formation in evolving interfaces]{Generic equations for pattern formation in evolving interfaces}

\author{Mario Castro}
\address{Grupo Interdisciplinar de Sistemas Complejos (GISC)
and Grupo de Din\'amica No Lineal (DNL), Escuela T\'ecnica Superior
de Ingenier{\'\i}a (ICAI), Universidad Pontificia Comillas,
E-28015 Madrid, Spain}
\author{Javier Mu\~noz-Garc\'{\i}a}
\address{Departamento de Matem\'aticas and GISC,
Universidad Carlos III de Madrid, 
E-28911 Legan\'es, Spain}
\author{Rodolfo Cuerno}
\address{Departamento de Matem\'aticas and GISC,
Universidad Carlos III de Madrid, 
E-28911 Legan\'es, Spain}
\author{M.\ Garc\'{\i}a-Hern\'andez}
\address{Instituto de Ciencia de Materiales de Madrid (CSIC),
E-28049 Madrid, Spain}
\author{L.\  V\'azquez}
\address{Instituto de Ciencia de Materiales de Madrid (CSIC),
E-28049 Madrid, Spain}

\begin{abstract}
We present a general formalism to derive the evolution equations describing
1D and isotropic 2D interface-like systems, based on symmetries, conservation
laws, multiple scale arguments, and exploiting the relevance of coarsening
dynamics.
Our approach becomes specially significant in the presence of surface
morphological instabilities and allow us to classify the most relevant
nonlinear terms in the continuum description of these systems. The theory
applies to systems ranging from eroded nanostructures to macroscopic pattern
formation. In particular, we show the validity of the theory for novel
experiments on ion plasma erosion.
\end{abstract}

\maketitle

\section{Introduction}

Many surfaces and interfaces, which are very different in nature and that occur
at very different lengthscales, produce interfaces which are startlingly
similar. This similarity originates in the competition between stabilizing and
destabilizing physical mechanisms, that are insensitive to the specific value
of the interface height $h(x,t)$  (this function providing the height of a
interface above position $x$ on a 1D substrate, at time $t$). Namely, such
mechanisms are invariant under global height translations of the form $h(x,t)
\to h(x,t) + \xi$, with $\xi$ an arbitrary constant (shift
simmetry~\cite{hentschel}). For instance, interesting [(sub)micrometric]
surface features develop by growth or erosion using various techniques, such as
electrochemical deposition (ECD)~\cite{ecd}, chemical vapor deposition
(CVD)~\cite{cvd}, ion beam sputtering (IBS)~\cite{valbusa} or molecular beam
epitaxy (MBE)~\cite{politi}, but remarkably also in macroscopic systems, such
as aeolian sand dunes~\cite{valance} or underwater (vortex) ripples in
sand~\cite{krug}.

For cases in which dynamical instabilities are irrelevant or absent, the
shift symmetry has been argued to lead, in the presence of noise, to scale
invariant surface morphologies. This is the {\em generic scale invariance}
\cite{grinstein} associated with the universality classes~\cite{barabasi}
of kinetic roughening. However, the situation differs in the presence of
morphological instabilities, that are the landmark of pattern formation.
In these cases, if the height field is seen as a measure of the {\em
amplitude} of perturbations around a reference homogeneous state, the
conservation law associated with the shift symmetry leads to a large-scale
instability \cite{nep}. This might prevent a unified description by means of
a universal equation, such as the Ginzburg-Landau equation is for the case
of short-wavelength instabilities \cite{nep,ch}. Thus, although the systems
mentioned above are seemingly governed by similar equations, a general
theory of evolving interfaces in which patterns arise is still lacking. In
particular, it seems that for every experimental system a specific theory
needs to be developed.

In this paper, we present a general formalism for pattern formation in 1D
and isotropic 2D surfaces, that relates with previous approaches in the theory
of dynamic scaling of rough interfaces \cite{hentschel,csahok,prl05},
exploiting systematically the formulation of the interface evolution within the
assumption of a (generalized) relaxational dynamics. Through symmetries
and multiple scale arguments, our approach helps to understand why some
effective equations presenting a morphological instability appear almost
ubiquitously in the experimental systems mentioned above. Our method enables
classification of the most relevant nonlinear terms for these systems, and
stresses the relevance of the phenomenon of coarsening to their continuum
description. We have also performed some experiments on ion plasma erosion to
explore non-trivial implications of our theory. Moreover, we have applied it to
other relevant and diverse fields, ranging from thin film production to surface
patterns in macroscopic systems, such as fluid waves or some granular systems.

In principle, we consider that the evolution of $h$ is local and,
mathematically, takes the form $\partial_t h={\cal G}(\{h\})+\eta$, where
$\{h\}$ stands for $h$ and its spatial derivatives, and $\eta$ is a noise term
that we will neglect hereafter. Previous phenomenological proposals for the
functional $\cal G$ have been built upon geometry arguments~\cite{csahok},
or upon symmetry and conservation requirements~\cite{barabasi}. In the first
case, the evolution is governed by the local curvature and its derivatives
along the interface. This excludes contributions to $\cal G$ that are due
to, e.g., anisotropies \cite{golovin}. Likewise, the second approach can be
misleading. For instance, if one considers non-conserved interface dynamics
(namely, if $\cal G$ cannot be derived from a current) under the shift
symmetry, one could naively drop terms like $\frac12\partial_x^2h^2$ (because
it can be written as $\partial_xj$) or $h\partial^2_xh$ (because it is not
shift invariant). However, $\cal G$ could contain a term proportional to
$\p^2$, namely, the sum of both terms. This maybe innocuous in the presence
of generic scale invariance, but not in the presence of instabilities, hence a
systematic method to classify the terms in the equation is needed.

\section{Theory}

\subsection{Non-conserved dynamics}
We proceed on by studying separately systems with non-conserved and conserved
dynamics. As in Refs.\ \cite{hentschel,prl05}, we assume that the dynamics of
the systems is relaxational in a generalized sense  (namely, with a
non-constant mobility). Although this assumption is not strictly necessary, it
clarifies greatly the interpretation of the coefficients of the non-linear
terms below. Hence,
\begin{equation}
\partial_t h= -\Gamma(\{h\})\frac{\delta F(\{h\})}{\delta h},
\label{main-nc}
\end{equation}
where $\Gamma$ ($\geq 0$) and $F$ are, respectively, a function
and a functional of the height field and its spatial derivatives.
Hereafter, we restrict $\{h\}$ to be $\{h, \partial_xh,
\partial_x^2h,\partial_x^3h\}$~\cite{note_deriv}.

If the relaxation rate $\Gamma$ is a constant, symmetry under global shifts
forbids any explicit dependence of $F$ on $h$, otherwise, what needs to be
independent of $h$ is the right hand side of Eq.\ (\ref{main-nc}).
Consequently, one can write (note our difference in sign convention from
\cite{hentschel}):
\begin{eqnarray}
{\mathcal F}(h,\partial_xh,\partial_x^2h,\partial_x^3h) & = & e^{-s h}
{\mathcal F}_1(\partial_xh,\partial_x^2h,\partial_x^3h), \label{redefF} \\
\Gamma(h,\partial_xh,\partial_x^2h,\partial_x^3h) & = & e^{s h}
\Gamma_1(\partial_xh,\partial_x^2h,\partial_x^3h),
\label{redefG}
\end{eqnarray}
where the height scale $s$ is the generator of the group of symmetry under
translations in $h$, and we have written $F(\{h\}) = \int {\mathcal F}(\{h\})
\, dx$.

Close to the instability threshold, we can assume that local slopes are small,
$|\partial_xh|\ll 1$, and we can expand $\Gamma_1$ and ${\mathcal F}_1$ in
power series. Therefore,
\begin{eqnarray}
\Gamma_1 & = & \sum_{i,j,k}\gamma^{(ijk)}(\partial_xh)^i(\partial_x^2h)^j
(\partial_x^3h)^k,\label{taylorG} \\
{\mathcal F}_1 & = & \sum_{i,j,k}f^{(ijk)}(\partial_xh)^i(\partial_x^2h)^j
(\partial_x^3h)^k.\label{taylorF}
\end{eqnarray}
Using these expansions in Eq.\ (\ref{main-nc}), we obtain a list of terms
compatible with the shift symmetry. The resulting equations are quite
involved. Hence, we first consider the linear terms. Later we will include the
most relevant non-linear terms. Thus, we find
\begin{equation}
\partial_th=V_\perp+V_\parallel\partial_xh+\nu \partial_x^2h+
\beta \partial_x^3h-K\partial_x^4h+\ldots,
\label{linear}
\end{equation}
where
\begin{eqnarray}
V_\perp&=&s \gamma f , \quad V_\parallel=s \gamma^{(100)}f ,\\
\nu&=&s \gamma^{(010)}f +2s \gamma f^{(010)}+2 \gamma f^{(200)},\label{nu}\\
\beta&=&s\gamma^{(001)}f , \quad
K=-2\gamma f^{(020)}+2\gamma f^{(101)}, \label{Keq}
\end{eqnarray}
and $\gamma\equiv\gamma^{(000)}$, $f\equiv f^{(000)}$. The real part of the
dispersion relation for single mode solutions of Eq.\ (\ref{linear}),
$h(x,t)\sim \exp[\omega_qt+ikx]$, reads $\Re e(\omega_q)=-\nu q^2-Kq^4$;
consequently, a morphological instability arises if $\nu<0$ and $K>0$. The
characteristic length-scale for the ensuing pattern will be roughly given by
the wavelength of the mode maximizing $\omega_q$, namely,
$l=2\pi\sqrt{2K/|\nu|}$. When nonlinear terms are incorporated into Eq.\
(\ref{linear}), this length-scale can grow in a {\em coarsening} process
\cite{politimisbah}, that will help us to identify which nonlinear
contributions are actually playing a role in the full  dynamics. Close to the
instability threshold $\nu=0$, Eq.\ (\ref{nu}) can be written as $\nu= -2\gamma
f^{(200)}(s-s_c)/s_c\equiv -2\gamma f^{(200)}\epsilon$, with $\epsilon$ a small
dimensionless parameter, and $s_c=-2\gamma f^{(200)}/(\gamma^{(010)}f+2\gamma
f^{(010)})$. Note the length-scale $l$ diverges as $\epsilon^{-1/2}$ in the
limit $\epsilon\to 0$, as $K$ is $s$-independent, according to Eq.\
(\ref{Keq}). Consequently, we can rescale length, time and height as
$x\rightarrow x/l$, $t\rightarrow t/l^z$ and $h\rightarrow h/l^\alpha$, and,
using the multiple scales method, look for the most relevant nonlinear term.
 Similarly to the theory of dynamic
scaling \cite{barabasi}, we obtain it to be $\lambda(s,s^2) \p^2$, with $
\lambda(s,s^2)\equiv s \mF\G^{(200)} - s^2\G \mF^{ (010)} -s\G \mF^{(200)},$
where $(s,s^2)$ is used to stress the polynomial dependence of $\lambda$ on
$s$. For many experimental situations, however, the hydrodynamic limit may lie
beyond practical reach \cite{cuernocastro,cv}, in which case the final
structure depends crucially on subdominant terms, that must be included in
the evolution equation in order to capture essential features of the original
nonlinear system. This can be done within our multiple scale framework. 
For instance, the ratio between $\p^2$ and $\p\pd$ scales as:
\begin{displaymath}
\frac{\p^2}{\p\pd}\sim l
\end{displaymath}
so, in the limit $l\rightarrow \infty$, the first one
dominates the second one. This calculation be reproduced for every term 
resulting in the series expansion, with the results that
follow for the first correcting terms, in order of relevance:
\begin{eqnarray}
&&\lambda_{110}(1,s,s^2)\p\pd, \quad
\lambda_{300}(1,s,s^2)\p^3, \nonumber \\
&&\lambda_{2}(1,s)\partial_x^2\p^2, \quad \lambda_{101}(s)\p\ppt, \label{nlr}\\
&&\lambda_{210}(1,s,s^2)\p^2\pd, \quad
\lambda_{400}(s,s^2)\p^4, \nonumber
\end{eqnarray}
where
\begin{eqnarray}
\lambda_{110}&=& s \mF\G^{(1,1,0)}+2s
\G^{(1,0,0)}\mF^{(0,1,0)}+2\G^{(1,0,0)}\mF^{(2,0,0)}
+3s \G\mF^{(1,1,0)}\nonumber \\
&&+3s^2\G\mF^{(0,0,1)}+6\G\mF^{(3,0,0)},\\
\lambda_{300}&=& s(-s\G^{(1,0,0)}\mF^{(0,1,0)}-\G^{(1,0,0)}\mF^{(2,0,0)}
-s^2\G\mF^{(0,0,1)}-s\G\mF^{(1,1,0)}\nonumber\\
&&-2\G\mF^{(3,0,0)} +\G^{(3,0,0)}\mF),\\
\lambda_{101}&=&
5s\G\mF^{(0,2,0)}/2-5s\G\mF^{(1,0,1)}/2+s\G^{(1,0,1)}\mF-s\G^{(0,2,0)}f/2-s\G^{(0,1,0)}\mF^{(0,1,0)} \nonumber \\
&&-\G^{(0,1,0)}\mF^{(2,0,0)}\\
\lambda_{2}&=&
s\G\mF^{(0,2,0)}/2-3s\G\mF^{(1,0,1)}/2+3s\G^{(0,2,0)}f/2+s\G^{(0,1,0)}\mF^{(0,1,0)}+\G^{(0,1,0)}\mF^{(2,0,0)}
\nonumber \\
\lambda_{210}&=&
-2s^2[\mF^{(0,1,0)}\G^{(0,1,0)}-3\mF^{(0,0,1)}\G^{(1,0,0)}+\G(2\mF^{(0,2,0)}
-6\mF^{(1,0,1)})] \nonumber \\
&&+4\mF^{(2,0,0)}\G^{(2,0,0)}+
s(6\G^{(1,0,0)}\mF^{(1,1,0)}-2\G^{(0,1,0)}\mF^{(2,0,0)}\nonumber \\
&&+4\mF^{(0,1,0)}\G^{(2,0,0)} +8\G\mF^{(2,1,0)}+2\mF\G^{(2,1,0)})
+12\G^{(1,0,0)}\mF^{(3,0,0)}\nonumber \\
&&+24\G\mF^{(4,0,0)} ,\\
\lambda_{400}&=& s(-s\G^{(2,0,0)}\mF^{(0,1,0)}-\G^{(2,0,0)}\mF^{(2,0,0)}
-s^2\G^{(1,0,0)}\mF^{(0,0,1)}\nonumber \\
&&-s\G^{(1,0,0)}\mF^{(1,1,0)}
-2\G^{(1,0,0)}\mF^{(3,0,0)} -s^2\G\mF^{(1,0,1)}-s\G\mF^{(2,1,0)}\nonumber \\
&&-3\G\mF^{(4,0,0)} +\mF\G^{(4,0,0)}),
\end{eqnarray}
and we have used that $(\partial_x^2 h)^2 = \big[\partial_x^2(\partial_x h)^2 -
\partial_x h \partial_x^3 h\big]/2$.

An example is provided by the so-called dissipation-modified Korteweg-de Vries equation,
\begin{equation}
\partial_th=\nu \partial_x^2h+\beta \partial_x^3h-K\partial_x^4h+
\lambda (\partial_x h)^2 + \lambda_{110}\partial_{x}^2h\partial_x h ,
\label{dmKdV}
\end{equation}
proposed \cite{bn} as the generic amplitude equation for weakly non-linear
waves, appearing in contexts from Marangoni or Rayleigh convection to periodic
waves in binary fluid mixtures (see Refs.\ in \cite{bn}). In our surface
context, it describes e.g.\ dynamics of (1D) ion-sputtered interfaces under
oblique incidence, see Refs.\ in \cite{mcb}. In general, while the term $\p^2$
dominates asymptotically, it interrupts coarsening \cite{us05,mcc}. This is
actually the case for Eq.\ (\ref{dmKdV}). Consequently, eventual coarsening of
the pattern must be due to other terms in Eqs.\ (\ref{nlr}). E.g., if one adds
the $\lambda_2$ term in Eq.\ (\ref{nlr}) to the right hand side of Eq.\
(\ref{dmKdV}), (interrupted) coarsening indeed occurs \cite{mcc2}. Further
details and examples are provided below in Section \ref{subsec3.1}.

\subsection{Conserved dynamics}
\label{cons-dyn}
In the case of conserved dynamics for the interface, we can write~\cite{hentschel}
\begin{equation}
\partial_th=\partial_x\left(\Gamma(\{h\}) \,
\partial_x\frac{\delta F(\{h\})}{\delta h} - j_{\rm nr} \right).
\label{cons}
\end{equation}
The assumption of  (generalized) relaxational dynamics is now stronger
than in the non-conserved one, the operator between $\Gamma$ and $\delta
F/\delta h$ constraining the form of the possible terms in a multiple scale
expansion (see below). In order to make the problem as general as possible, we
need to include the non-relaxational current, $j_{\rm nr}$, into the right hand
side of Eq.\ (\ref{cons}). The origin of this type of current has been widely
discussed e.g.\ in the context of mound formation in homoepitaxial
growth~\cite{krug2,politi}. Assuming $j_{\rm nr}$ to depend only on derivatives
of $h$ in the form
\begin{equation}
j_{\rm nr}= \sum_{i,j,k}J^{(ijk)}(\partial_xh)^i(\partial_x^2h)^j(\partial_x^3h)^k,
\label{taylorJ}
\end{equation}
and using Eqs.\ (\ref{taylorG})-(\ref{taylorF}) as above, Eq.\ (\ref{cons})
takes the form
\begin{eqnarray}
\partial_th & = & \nu^c\partial_x^2h+\beta^c \partial_x^3h-K^c\partial_x^4h+
\lambda_{110}^c\p\pd\nonumber \\
 & + & \lambda_2^c\partial_x^2\p^2+\lambda_{210}^c\p^2\pd,
\label{sand_eq}
\end{eqnarray}
where all the coefficients in Eq.\ (\ref{sand_eq}) depend on $\gamma^{(ijk)}$,
$f^{(ijk)}$ and $J^{(ijk)}$. Thus,
\begin{eqnarray}
\nu^c&=&s^2\G\mF-J^{(1,0,0)}\\
\beta^c&=&-J^{(0,1,0)}\\
K^c&=&J^{(0,0,1)}+2s\G\mF^{(0,1,0)}+\G\mF^{(2,0,0)}\\
\lambda_{110}^c&=&2s^2\gamma^{(1,0,0)}f-J^{(2,0,0)}\\
\lambda_2^c&=& s(s\G^{(0,1,0)}\mF+4s\G\mF^{(0,1,0)}+2\G\mF^{(2,0,0)})/2\\
\lambda_{210}^c&=& s^2(-3s\G\mF^{(0,1,0)}-3\G\mF^{(2,0,0)}/2 +3\G^{(2,0,0)}\mF/2)
\end{eqnarray}

An interesting result derived from the above equations is that, whenever
$j_{\rm nr}=0$, the condition for unstable growth reads $\epsilon\propto
s^2$ so that power counting (with $l\sim \epsilon^{-1/2}$, $t\sim l^z$ and
$h\sim l^{\alpha}$) provides terms which give a divergent scaling of the
local slope, namely, $|\partial_xh|$ scales with a positive power of $l$. 
This means that the small slope
approximation breaks down, and the expansions above are no longer valid,
in analogy e.g.\ with the so-called superrough scaling in the context of
surface kinetic roughening~\cite{barabasi}, occurring specifically in the
case of conserved dynamics. In these conditions, detailed knowledge of system
specifics is needed and a strongly nonlinear analysis must be performed in
order to obtain the correct interface equation, as in~\cite{pierre_gillet}.

\section{Results}
Thus far, we have derived the most general and relevant interface equations
compatible with the shift symmetry. We proceed on by testing the validity of
the theory and, more importantly, its capability to identify the right terms of
the evolution equation in each context, based on experimental or theoretical
evidences, through our procedure that accounts for conservation,
symmetries and coarsening, in systems where a surface instability is present.
To illustrate  this method of analysis, we discuss several 1D
examples of systems with lengthscales ranging from microscopic to macroscopic,
for each type of dynamics, both non-conserved and conserved. Moreover, some of
the examples correspon to isotropic 2D surfaces.

\subsection{Application to one-dimensional systems} \label{subsec3.1}

To begin with, we consider non-conserved growth of nanometric sized
patterns by Ion Beam Sputtering (IBS). In this technique, the surface of a
solid target is bombarded isotropically with a flux of energetic ions.
Experimentally, a typical length scale develops at early times, that is seen to
grow later (coarsening). The system is not conserved due to the mass loss as a
consequence of the bombarding process. Finally, the experimentally obtained
surfaces do not develop steep slopes (so a small slope approximation is well
justified). With these ingredients in mind, our theory predicts that the
evolution equation for these systems must contain the term proportional to
$\lambda$ due to the non-conserved nature of the experiment; and terms
$\lambda_2$, $\lambda_{101}$ and $\lambda_{210}$ due to the presence of
coarsening. Hence,
\begin{eqnarray}
\partial_th&=&\nu\pd -K\pc +\lambda\p^2+\lambda_2\partial_x^2\pd+
\lambda_{101}\p\ppt\nonumber \\
&&+\lambda_{210}\p^2\pd \label{ibs_eq}
\end{eqnarray}
Consistently with the above expectation, a hydrodynamic theory of erosion
has been reported~\cite{us05} which couples a fast field describing the
density of mobile particles at the surface, with a slow one which
corresponds to the observed experimental surface. This theory allows to relate
experimental magnitudes with corresponding equation parameters and leads
to, precisely, Eq.\ (\ref{ibs_eq}).

At intermediate length scales (microns to millimeters), we can take
growth by ECD~\cite{ecd}, where the surface of an aggregate grows by
incorporation of cations from a solution. The assumption of locality above is
reasonable for a small sticking probability, i.e., if the cations do not attach
to the first visited surface site but are, rather, allowed to diffuse further
until aggregation occurs. The cation flux at any surface point is isotropic,
hence the system is reflection ($x\rightarrow -x$) symmetric. Finally, the
characteristic length of the unstable structures observed experimentally does
not grow in time, i.e., there is no coarsening. These data suggest that surface
dynamics for ECD is described by the Kuramoto-Sivashinsky (KS) equation,
namely, $V_\parallel=\beta=0$, due to the reflection symmetry, and that all the
terms in Eqs.\ (\ref{nlr}) are negligible, compatible with the absence of
coarsening. The same scenario is observed in surface growth by CVD~\cite{cvd}.
Indeed, the KS equation has been derived explicitly for both ECD and CVD from
constitutive equations, in agreement with our results~\cite{cuernocastro}.

Our final example of non-conserved interface growth is taken from surface
instabilities in fluids due to surface tension gradients (see \cite{colinet}
for a comprehensive reference), an standard example being provided e.g.\ by
Marangini-B\'enard convection \cite{garazo-velarde}. In this system, a thin
fluid layer is subject to an external transverse temperature gradient, there
being non-conserved fluxes that implement the dissipation in the system. When
heating from below a convective instability sets in whose spatial
characteristics do not coarsen in time. This instability leads to travelling
waves on the free surface of the system that break the $x\rightarrow -x$
reflection symmetry. An asymptotic study of the corresponding hydrodynamic
problem leads to the equation \cite{garazo-velarde,colinet}
\begin{equation}
\partial_th=\nu\pd -K\pc +\lambda\p^2+\lambda_{110}\p\pd,
\end{equation}
in agreement with the general expectation from our approach under the
constraints considered.

As an important example of conserved interface dynamics at nanometric
scales, we consider the growth by Molecular Beam Epitaxy (MBE) of
surfaces that are vicinal to a high symmetry crystalline
orientation~\cite{michely_krug}. In this type of experiments, the molecules
incorporated from the chamber arrive isotropically at the surface (and hence
reflection symmetry is fullfilled), the relevant one-dimensional
interface being given by the location of the step separating one terrace
from an adjacent one. Under conditions in which the rate of adatom
evaporation from the  terrace is vanishingly small, the dynamics is
conserved. It is precisely in this case that the small slope approximation
has been seen to break down \cite{pierre_gillet}, as for Eq.\ (\ref{cons}) for
$j_{\rm nr} = 0$. However, if adatom desorption occurs on the terraces, the
system becomes non-conserved and the relevant interfacial description is
provided by the Kuramoto-Sivashinsky equation \cite{karma_misbah}, as expected
from our above arguments.

Finally, an example of conserved interface growth in a macroscopic system
is provided by the formation of macroscopic ripples in aeolian sand
dunes~\cite{csahok}. In this case the wind direction breaks the reflection
symmetry and, surface dynamics at the sand bed being conserved due to the
conservation in the number of sand grains, one would expect Eq.\
(\ref{sand_eq}) to hold as the relevant continuum equation. Indeed, this has
been shown to be the case within the so-called hydrodynamic approach to pattern
formation in these systems~\cite{csahok}.

The results in this section, as well as some other experimental systems
that can be analyzed along similar lines, are summarized in Table~\ref{table},
where we provide references to the experimental observations and to theoretical
derivations compatible with our conclusions. We provide the reader with broader
references when connection with specific experiments is less unambiguous.

\begin{center}
\begin{table}[!h]
\begin{tabular}{l c c c c c c c}\hline\hline
System & Type & RS & Coars. & Nonlinearities & Exp. &
Theo.  \\\hline
ECD/CVD & NC& Yes &No & $\lambda$ & \cite{ecd_exp} & \cite{cuernocastro} \\
IBS& NC& Yes & Yes & $\lambda$, $\lambda_{2,101,210}$ &
\cite{facsko} & \cite{us05} \\
Thin films & NC& Yes & Yes & $\lambda$, $\lambda_2$ & \cite{mayr} & \cite{raible} \\
Fluid waves & NC & No & No & $\lambda$, $\lambda_{110}$ & \cite{colinet} &
\cite{colinet} \\ \hline
MBE (singular) & C & Yes & Yes & $\lambda_{210}^c$ & \cite{michely_krug} & \cite{politi}
\\
MBE (vicinal) & C & Yes & No & SN & \cite{michely_krug}& \cite{pierre_gillet} \\
Sand ripples & C & No & Yes & $\lambda^c_{2,110,210}$ &
\cite{rioual}& \cite{valance} \\
Vortex ripples & C & Yes & Yes & SN & \cite{vortex_exp} & \cite{krug} \\
\hline \hline
\end{tabular}
\caption{Physical examples. Non-conserved (NC), conserved (C), strongly-nonlinear (SN).
RS means reflection symmetry ($x\to -x$), and ``Coars.'' is coarsening.}
\label{table}
\end{table}
\end{center}

\subsection{Application to novel experiments}
\begin{figure}[!ht]
\begin{center}
\includegraphics[width=\textwidth,clip=]{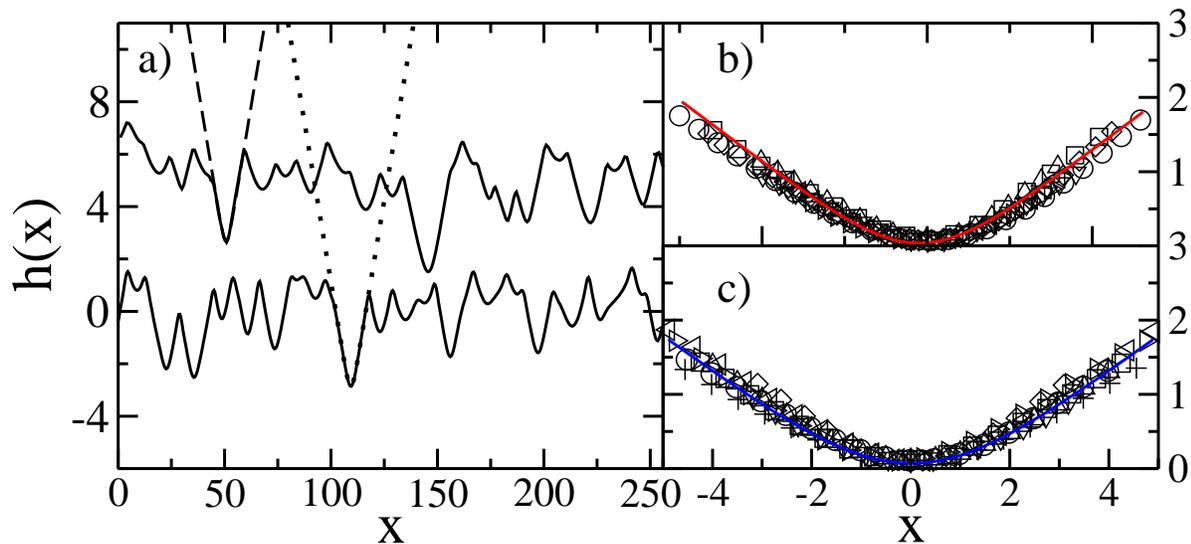}
\end{center}
\caption{a) Cross section of Fig.\ \ref{fig2}a  morphologies, both numerical (lower curve) obtained with $\nu=1$, $K=0.11$,
$\lambda_{210}=0.052$ and $\lambda=-0.05$ and
experimental (upper curve). Dashed line stand for a fit of a the
experimental cell shape
to Eq.\ (\ref{shape}), which allow to obtain the parameters used in the
numerical simulations. Dotted line is a translation of the dashed line to
stress the self-consistency of the parameter extraction. b) Fit to Eq.\ (\ref{shape}) of 7 {\em valleys} extracted from top curve in Fig.\ \ref{fig1}a. Red
solid line stand for a dashed and dotted line in Fig.\ \ref{fig1}a. c)
Fit to Eq.\ (\ref{shape}) of 8 {\em valleys} extracted from bottom curve
in Fig.\ \ref{fig1}a. Blue solid line  stand for a dashed and dotted line in Fig.\ \ref{fig1}a.}
\label{fig1}
\end{figure}

\begin{figure}[!ht]
\begin{center}
\includegraphics[width=\textwidth,clip=]{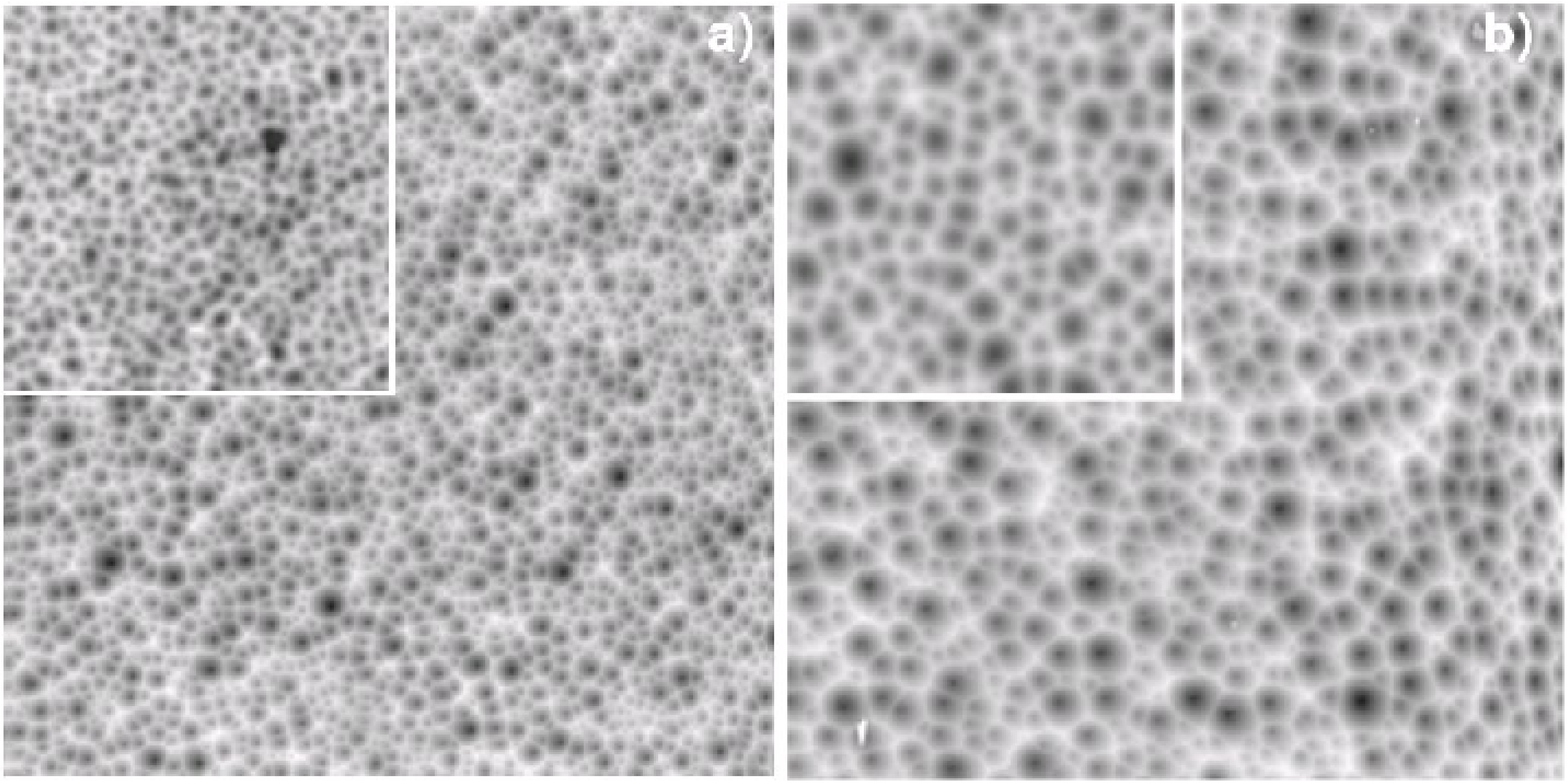}
\end{center}
\caption{a) Ion plasma eroded Si(100) at $t_e=3$ h. Inset: Numerical simulation
of Eq.\ (\ref{cells_eq_final}) at $t_s=35$ with $\nu=1$, $K=0.11$,
$\lambda_{210}=0.052$ and $\lambda=-0.05$. The lateral size of the image is
$50$ $\mu$m. b) Ion plasma eroded Si(100) at $t_e=6$ h. Inset: Simulation as in
a) for $t_s = 70$. The lateral size of the image is $50$ $\mu$m. } \label{fig2}
\end{figure}

The examples in last subsection show how, based on general considerations, one
can construct a general equation in good agreement with current specific
theories. Notwithstanding, we want to emphasize the wide range of generality of
our method by applying it to novel experiments on ion plasma erosion. Thus, a
Si(100) wafer (2 inches diameter) was immersed in an argon plasma, which was
confined magnetically, at a pressure of $5\times 10^{-3}$ mbar. The only
parameter that was changed in the experiments was the immersion (i.e.\
sputtering) time. After the sputtering process, the central part of the wafer
was analyzed by Atomic Force Microscopy operating in tapping mode with a
silicon cantilever. Under these conditions the flux of incoming ions on the
central part of the wafer is mainly isotropic. We plot two snapshots of the
surface taken at 3h (Fig.\ \ref{fig2}a) and 6h (Fig.\ \ref{fig2}b). Moreover,
we have measured a  non-zero coarsening exponent $n=0.53\pm 0.02$ in $\ell
\sim t^n$, $\ell$ being the mean cell diameter. The system being isotropic, we
can generalize our results above to this case.  Out of the most relevant
subdominant terms in Eq.\ (\ref{nlr}) (once generalized to two dimensions)
compatible with isotropy, $\lambda_2$ \cite{raible,mcc} and $\lambda_{210}$
\cite{emmott,golovin2} are known in the literature to induce coarsening
behavior when taken each of them in combination with $\lambda_{100}$ and a KS
dispersion relation. Extensive numerical simulations \cite{mcc2} show 
that the term $\lambda_{101}$ also induces coarsening, but the term
$\lambda_{400}$ does not.
Hence, thus far we are left with an equation of the following form:
\begin{eqnarray}
 \partial_t h&=& - \nu \nabla^2 h- K \nabla^4 h +\lambda (\nabla h)^2
 + \lambda_2 \nabla^2(\nabla h)^2 \nonumber\\
 && + \lambda_{101} \nabla h \cdot \nabla (\nabla^2 h)
 + \lambda_{210} (\nabla h)^2(\nabla^2 h).
 \label{cells_eq}
\end{eqnarray}
>From the experimental data we can extract some quantitative information about
the coefficients in Eq.\ (\ref{cells_eq}). Thus, a closer inspection of the
experimental surface cross section (Fig.\ \ref{fig1}) reveals that {\em cell}
shape can be approximated by
\begin{equation}
(\nu^2/K\lambda_{210})^{1/2}\log(\cosh[(2K/\nu)^{1/2}x]),
\label{shape}
\end{equation}
which is a solution of Eq.\ (\ref{cells_eq}) in 1D with
$\lambda_2=\lambda_{101}=0$ (and, consequently, approximately valid for large
cells in the isotropic 2D case). This suggests that $\lambda_2$ and
$\lambda_{101}$ are negligible since large values for them would lead, rather,
to parabolic cells~\cite{us05}.

Consequently, Eq.\ (\ref{cells_eq}) reduces effectively to
\begin{equation}
 \partial_t h = - \nu \nabla^2 h- K \nabla^4 h +\lambda (\nabla h)^2
 + \lambda_{101} \nabla h \cdot \nabla (\nabla^2 h), 
  \label{cells_eq_final}
\end{equation}
the so-called convective Cahn-Hilliard equation~\cite{golovin2}. For short
times (linear regime), the typical lengthscale allows determination of the
$K/\nu$ value, while remaining parameters are estimated by trial and error in
order to improve the resemblance with the experimental morphologies. Results of
such numerical simulations are compared with experiments in Fig.\ \ref{fig2}
showing an excellent agreement between them. Thus, the small panels (upper
insets) display two snapshots of the numerical integration of Eq.\
(\ref{cells_eq_final}) and the large panels the equivalent results from
experiments.


\section{Discussion and conclusions}
Although table I is not meant to be exhaustive, it leads to some general
observations. Thus, as anticipated above, the occurrence of coarsening is an
important property in guiding the identification of the appropriate nonlinear
terms of a given system. This phenomenon was associated with the seeming
inability to describe patterns that are stable both in wavelength and amplitude
by means of local height equations \cite{krug,politimisbah}. However, there may
exist now counterexamples \cite{us05,mcc} for this shortcoming,  at least
for an order range that is much larger than the single cell size. As a
consequence, the task of determining the necessary and sufficient conditions
for coarsening \cite{politimisbah}, as well as the possible values
(universality classes) for the coarsening exponent $n$, remains to be
completed, lying outside the scope of this paper. Note that, once coarsening
occurs, the fact that it {\em interrupts} \cite{politimisbah} or not, and the
value of exponent $n$, both depend on the type of leading and subleading
nonlinearities that appear in the evolution equation. Hence, these provide
additional criteria in order to propose a continuum description for a given
physical system,  although it can nevertheless be the case that two
different subleading terms lead to the same coarsening exponent value, such as
is the case for $\lambda_2$ and $\lambda_{210}$, both of them yielding $n=1/2$
\cite{raible2, emmott}. Moreover, system dimensionality and the (possibly
related) relevance of noise may influence the value of the coarsening exponent
in the corresponding universality classes.

In summary, we have provided a systematic method to obtain and classify the
relevant nonlinear terms in interfacial systems which present
morphological instabilities and the shift symmetry. Besides, we have
applied our method to some reported systems in the literature as well as to
novel experiments on ion plasma erosion performed for this purpose.

\ack This work has been partially supported by MECD (Spain), through Grants
Nos.\ BFM2003-07749-C05-01, BFM2003-07749-C05-02 and BFM2003-07749-C05-05,
FIS2006-12253-C06-01, FIS2006-12253-C06-03, FIS2006-12253-C06-06, and the FPU
programme (J.M.-G.).


\begin{thebibliography}{99}

\bibitem{hentschel} H.\ G.\ E.\ Hentschel, J. Phys. A: Math. Gen {\bf 27},
2269 (1994).

\bibitem{ecd} W. Schwarzacher, J. Phys. Condens. Matter {\bf 16}, R859 (2004).

\bibitem{cvd} F. Jensen and W. Kern, in {\em Thin Film Processes II},
edited by J. L. Vossen and W. Kern (Academic, Boston, 1991).

\bibitem{valbusa} U. Valbusa, C. Boragno, and F. Buatier de Mongeot, J. Phys.
Condens. Matter {\bf 14}, R8153 (2002).

\bibitem{politi} P.\ Politi, G. Grenet, A. Marty, A. Ponchet, and J. Villain,
Phys. Rep. {\bf 324}, 271 (2000).

\bibitem{valance} A. Valance and F. Rioual, Eur. Phys. J. B {\bf 10}, 543
(1999).

\bibitem{krug} J.\ Krug, Adv. Complex Systems {\bf 4}, 353 (2001).

\bibitem{grinstein} G. Grinstein, in {\em Scale Invariance, Interfaces and
Non-Equilibrium Systems}, edited by A. McKane, M. Droz, J. Vannimenus, and D.
Wolf (Plenum Press, New York, 1995).

\bibitem{barabasi} A.-L. Barab\'asi and H. E. Stanley, {\em Fractal Concepts
in Surface Growth} (Cambridge University Press, Cambridge, 1995).

\bibitem{nep} A. A. Nepomnhyaschy, Physica D {\bf 86}, 90 (1995).

\bibitem{ch} M.\ C.\ Cross and P.\ C.\ Hohenberg, Rev. Mod. Phys. {\bf 65},
851 (1993).

\bibitem{csahok} Z.\ Csah\'ok, C.\ Misbah, and A.\ Valance, Physica D {\bf
128}, 87 (1999).

\bibitem{prl05} J.\ M.\ L\'opez, M.\ Castro, and R.\ Gallego, Phys. Rev.
Lett. {\bf 94}, 166103 (2005).

\bibitem{golovin} A.\ A.\ Golovin, S.\ H.\ Davis, and A.\ A.\ Nepomnyashchy,
Phys. Rev. E {\bf 59}, 803 (1999).

\bibitem{note_deriv} Dependence on higher order derivatives,
e.g. $\partial_x^4h$, only changes
$\beta$ and $K$ in Eq.\ (\ref{linear}), but modifies neither the
nonlinear terms nor the general discussion.

\bibitem{politimisbah} P.\ Politi and C.\ Misbah, Phys. Rev. Lett. {\bf 92},
090601 (2004); Phys. Rev. E {\bf 73}, 036133 (2006).

\bibitem{cuernocastro} R.\ Cuerno and M.\ Castro, Phys. Rev. Lett. {\bf 87},
236103 (2001).

\bibitem{cv} R. Cuerno and L. V\'azquez, in {\em Advances in Condensed Matter
and Statistical Physics}, edited by E. Korutcheva and R. Cuerno (Nova Science
Publishers, New York, 2004).


\bibitem{bn} D. E. Bar and A. A. Nepomnhyaschy, Physica D {\bf 86}, 586 (1995).

\bibitem{mcb} M. Makeev, R. Cuerno, and A.-L. Barab\'asi, Nucl. Instrum. Methods Phys.
Res., Sect. B 197, 185 (2002).

\bibitem{us05} M.\ Castro,
R.\ Cuerno, L.\ V\'azquez, and R.\ Gago, Phys. Rev. Lett. {\bf 94}, 016102
(2005); M.\ Castro and R.\ Cuerno. {\em ibid} {\bf 94}, 139601 (2005); J.
Mu\~noz-Garc\'ia, M. Castro, and R. Cuerno, {\em ibid.} {\bf 96}, 086101
(2006).

\bibitem{mcc} J. Mu\~noz-Garc\'ia, R. Cuerno, and M. Castro, Phys. Rev. E
(Rapid Commm.) (2006) in press.

\bibitem{mcc2} J. Mu\~noz-Garc\'ia, M. Castro, and R. Cuerno, unpublished.


\bibitem{krug2} J.\ Krug, M.\ Plischke, and M.\ Siegert, Phys. Rev. Lett. {\bf
70}, 3271 (1993).

\bibitem{pierre_gillet} O.\ Pierre-Louis, C. Misbah, Y. Saito, J. Krug, and P. Politi,
Phys. Rev. Lett. {\bf 80}, 4221 (1998); F.\ Gillet, O.\ Pierre-Louis, and C.\
Misbah, Eur. Phys. J. B {\bf 18}, 519 (2000).

\bibitem{colinet} P. Colinet, J. C. Legros, and M. G. Velarde, {\em Non-linear
Dynamics of Surface-Tension-Driven Instabilities} (Wiley-VCH, Berlin, 2001).

\bibitem{garazo-velarde} A. N. Garazo and M. G. Velarde, Phys. Fluids

\bibitem{michely_krug} T. Michely and J. Krug, {\em Islands, Mounds and Atoms}
(Springer, Berlin, 2004).


\bibitem{karma_misbah} A. Karma and C. Misbah, Phys. Rev. Lett. {\bf 71}, 3810
(1993).

\bibitem{raible} M. Raible, S. J. Linz, and P. H\"anggi, Phys. Rev. E {\bf 64},
031506 (2001), and references therein.

\bibitem{emmott} C. L. Emmott and A. J. Bray, Phys. Rev. E {\bf 54}, 4568
(1996).

\bibitem{golovin2} A.\ A.\ Golovin,
A.\ A.\ Nepomnyashchy, S.\ H.\ Davis, and M.\ A.\ Zaks,
Phys. Rev. Lett. {\bf 86}, 1550 (2001).

\bibitem{raible2} M. Raible, S. J. Linz, and P H\"anggi, Phys. Rev. E {\bf 62},
1691 (2000).

\bibitem{ecd_exp} P.\ L.\ Schilardi, O. Azzaroni, R. C. Salvarezza, and A. J. Arvia,
Phys. Rev. B {\bf 59}, 4638 (1999); C. L\'eger, J. Elezgaray, and F. Argoul,
Phys. Rev. E {\bf 61}, 5452 (2000).

\bibitem{facsko} S. Facsko, T. Dekorsy, C. Koerdt, C. Trappe, H. Kurz, A. Vogt,
and H. L. Hartnagel, Science {\bf 285} 1551 (1999); R.\ Gago, L. V\'azquez, R.
Cuerno, M. Varela, C. Ballesteros, and J. M. Albella, Appl. Phys. Lett. {\bf
78}, 3316 (2001).


\bibitem{mayr} S.\ G.\ Mayr, M.\ Moske, and K.\ Samwer, Phys. Rev. B {\bf 60},
16950 (1999).

\bibitem{rioual} See, e.g. in F.\ Rioual,
{\em Etude de quelques aspects du transport olien: processus de saltation et
formation des rides},
Ph. D. thesis, Universit\'e de Rennes 2002 (unpublished).

\bibitem{vortex_exp} A.\ Stegner and J.\ E.\ Wesfreid, Phys. Rev. E {\bf 60},
R3487 (1999).


\end{thebibliography}
\end{document}